\documentstyle[epsfig,preprint,aps]{revtex}
\begin{document}
\draft
\def\bra#1{{\langle #1{\left| \right.}}}
\def\ket#1{{{\left.\right|} #1\rangle}}
\def\bfgreek#1{ \mbox{\boldmath$#1$}}
\title{Medium dependence of the bag constant in the quark-meson coupling model}
%
\author{D. H. Lu, K. Tsushima, A. W. Thomas, A. G. Williams}
\address{Department of Physics and Mathematical Physics\break
 and
        Special Research Centre for the Subatomic Structure of Matter,\break
        University of Adelaide, Australia 5005}
\author{K. Saito}
\address{Physics Division, Tohoku College of Pharmacy,  
	Sendai 981, Japan}
\maketitle
\vspace{-9.3cm}
\hfill ADP-97-45/T273
\vspace{9.3cm}
\begin{abstract}
Possible variations of the quark-meson coupling (QMC) model are examined 
in which the bag constant decreases in the nuclear medium.
The reduction is supposed to depend on either the mean scalar field or 
the effective mass of the nucleon. It is shown that the electric and magnetic 
radii of the bound nucleon are almost linearly correlated with
the bag constant. Using the fact that the size of the bound 
nucleon inside a nucleus is strongly constrained by $y$-scaling data in 
quasielastic, electron-nucleus scattering, we set a limit for the reduction 
allowed in the bag constant for these two models.
The present study implies that the bag constant can decrease up to
10--17 \% at average nuclear density, depending on the details of the model. 
\end{abstract}
%

\section{Introduction}

Whether a nucleon bound in a nuclear medium (precisely speaking, 
a bound ``nucleon'' is a quasiparticle which has the same quantum numbers 
as those of the nucleon) would alter 
its properties from those of a free nucleon is of fundamental interest 
in nuclear physics\cite{QHD,DBHF,Guichon88,Yazaki90,Tony94,BR91}. 
It is a traditional approach that nucleonic (and mesonic) 
degrees of freedom play an essential role in describing nuclei, 
but it is also true that those nucleons themselves have substructure.
The idea of employing quark degrees of freedom to describe 
certain aspects of nuclear structure was motivated in part by the discovery of 
the EMC effect, namely, the discovery that the structure functions of 
the nucleon [as measured in deep inelastic scattering (DIS)] depend on the 
nucleus in which the nucleon is bound\cite{EMC}. 
It has been widely accepted that DIS is well described by quark and 
gluon dynamics, i.e., quantum chromodynamics (QCD).
Thus, it is reasonable to expect that quarks inside a nucleus may play 
important roles in some contexts, and their wave functions could be strongly
modified by the surrounding nuclear environment.

There is strong evidence for a reduction of the effective nucleon mass 
in the nuclear medium. In quantum hadrodynamics (QHD)\cite{QHD}
and the quark-meson coupling model (QMC)\cite{Guichon88,Yazaki90,Tony94}, 
the effective mass of nucleon at normal nuclear density,
in mean-field approximation (MFA), is respectively predicted to be roughly
40--50\% and 20--30\% smaller than the value in free space.  
The mechanism for this reduction is the interaction 
of the nucleon with the mean scalar field.
QCD sum rules independently lead to a similar prediction from an apparently 
different mechanism, i.e., the reduction of the quark condensate 
as the density increases\cite{QCDSR}.
In contrast, whether the size of the nucleon changes and by how much, 
are relatively less clear. The validity of Brown-Rho scaling\cite{BR91} 
for the nucleon radius could be  
much more questionable than it is for hadron masses.

The successes of QHD leave little doubt that relativistic nuclear 
phenomenology is essential in describing the bulk properties of nuclear matter
as well as the properties of finite nuclei.
A compelling feature of relativistic mean-field theory has been the
reproduction of spin-orbit splittings in finite nuclei  
and the description of spin observables in nucleon--nucleus scattering,  
after calibrating the model parameters by the equilibrium properties 
of symmetric nuclear matter (at normal nuclear density). 
These successful results could be ascribed to the reduction of the 
effective mass of the nucleon\cite{QHD}.

To incorporate the composite nature of the nucleon without losing
the successful phenomenology of QHD, Guichon proposed a
simple hybrid model\cite{Guichon88} (the quark-meson coupling (QMC) model) 
in which nuclear matter is described by non-overlapping, 
MIT bag nucleons\cite{MIT}. 
By analogy with QHD,  QMC describes the bulk properties 
of nuclear systems using scalar $(\sigma)$  and vector $(\omega)$ meson 
mean fields. 
However, the nucleon bound in nuclear medium here 
is no longer a point particle, 
it has  substructure -- quarks confined inside the nucleon bag.
It is the quark, rather than the nucleon itself, that is coupled to
the $\sigma$ and $\omega$ fields directly. 
As a result, the internal structure of the bound nucleon 
is modified by the surrounding medium with respect to that in free space.
The relative simplicity of the QMC model at the hadronic level, 
together with its phenomenological successes, makes it suitable for  many  
applications in nuclear physics\cite{Tony94,BM96,Aguirre97}.
There are similar models which use the harmonic oscillator\cite{BM96} 
or the color dielectric model\cite{NB93} 
instead of the MIT bag for the nucleon.

Within QMC, the small mass of the quark implies that the lower  
component of the quark wave function will be enhanced rapidly by the  
change of its environment (as the $\sigma$ field strength increases),
with a consequent decrease in the scalar baryon density. 
As the scalar baryon density itself is the source of the $\sigma$ field, 
this provides a mechanism for the saturation of nuclear matter 
where the quark substructure plays a vital role.
The extra degrees of freedom, corresponding to the internal structure of 
the nucleon, lead to a reasonable value for the nuclear incompressibilty, 
once the corresponding quark and meson coupling constants, 
$g^q_\sigma$ and $g^q_\omega$, are determined to reproduce 
the empirical values for the saturation density and binding energy 
of symmetric nuclear matter. One drawback in this approach is that the 
effective scalar and vector fields are significantly reduced, 
which in turn leads to somewhat insufficient spin-orbit interaction 
in finite nuclei (within MFA)\cite{Guichon96}.

With this difficulty in mind, Jin and Jennings postulated that 
the bag constant ($B$) might decrease along with a reduction of the 
effective nucleon mass ($m_N^*$)\cite{JJ96}. 
In their model, the bag constant had to be reduced dramatically
(approximately 35\% at normal nuclear density) 
to achieve ``the large and cancelling scalar and vector
potentials'' which arise in QHD and lead to the large nuclear 
spin-orbit forces. One undesirable side effect is 
that the nuclear incompressbility once again tends to be overpredicted 
(one of the original successes of Guichon's model was that 
it generated a reasonable value for the incompressibility). 
In addition, the nucleon size was found to increase unpleasantly 
because of the reduction of the bag constant. 
(For a recent study, see also Ref.\cite{Muller97}.)

On the other hand, there are already strong, experimental constraints
on the extent of such changes of nucleon size\cite{sick,Day90,coul}. 
Furthermore, 
new quasi-elastic $(e,e')$ and $(e,e'N)$ experiments at facilities 
such as Mainz, MIT-Bates and especially TJNAF
should improve these constraints in the near future.

In this paper we examine this issue, namely, how much the 
bag constant can be allowed to decrease in nuclear matter given those 
experimental constraints. For this purpose, we first 
solve self-consistently the equations of motion for the quarks  
and mesons in the modified QMC models\cite{JJ96}.
Then we apply the solutions obtained 
to calculate the electromagnetic form factors of the nucleon in medium.
From these calculated form factors, we extract the 
root-mean-square (r.m.s.) magnetic- and electric- radii. 
By comparing these with  the experimental data for the 
change of the nucleon size\cite{sick,Day90,coul,deForest83}
we can constrain the reduction of the bag constant.
For simplicity, possible off-shell effects of the bound nucleon 
in medium\cite{deForest83,Naus87} are  neglected in the present treatment, 
but should be added in future, more elaborate studies.
However, since each nucleon bag contains three valence 
quarks in the ground state which move independently
under the spherical cavity approximation to the MIT bag\cite{MIT}, 
the spurious center-of-mass motion has to be removed.
In this study, we use the Peierls-Thouless (PT) projection method
combined  with a Lorentz contraction for the nucleon internal 
wave function, as in Ref.~\cite{Lu97}. 
This technique for implementing the center-of-mass and
recoil corrections has been quite successful in the case of the 
electromagnetic form factors of nucleon in free space\cite{Lu97}. 

In conventional nuclear physics, the nucleon is immutable and 
the electromagnetic properties of a nucleus are often described by 
a combination of individual nucleon contributions and various  
meson exchange current (MEC) corrections.
Although we believe that the two body meson exchange 
is small in the present case, a full calculation of 
a nuclear form factor requires 
a consistent treatment of all MEC corrections. 
Since we will ignore the explicit MEC corrections here for simplicity, 
our results should be viewed as somewhat tentative with a  
need for future elaboration.

In section II, we briefly summarize our version of the modified QMC model
with the option for the reduction of the bag constant. As in Ref.~\cite{JJ96},
$B$ is related to either $\overline{\sigma}$ or $m_N^*$, however, 
the parameters are quite different because of the improved technical 
treatment of center-of-mass  motion.
In sec.~III, we first discuss the experimental status of the constraints
on the nucleon electromagnetic radii. The basics of the $y$-scaling analysis
are  presented. Then we show the formalism used to calculate the nucleon
electromagnetic form factors and r.m.s. radii. 
In sec.~IV we discuss  pionic corrections and the effects of meson-exchange 
currents. The numerical results are discussed in sec.~V and final conclusion
is given in sec.~VI.

\section{Aspects of the generalized QMC model}

We first review the standard treatment of the 
QMC model~\cite{Guichon88,Tony94,Guichon96}, 
which does not allow for any variation of the bag constant, $B$, 
in the nuclear medium.

In QMC, nuclear matter is described in terms of non-overlapping MIT bags,
and the quarks inside the bags interact directly with the scalar and vector
meson fields, where, the motion of the quarks is highly relativistic and 
typically they move much faster than the nucleon 
in the medium (the Fermi motion of the nucleon). Thus it is reasonable to 
assume that the quarks always have sufficient time to adjust their motion 
as the nucleon responds to the medium it is in so that they stay
in the lowest energy state\cite{Guichon96}.
In the instantaneous rest frame of a nucleon in symmetric nuclear matter 
(where the $\rho$ meson mean-field will vanish), the Lagrangian 
density for the QMC model may be given, 
\begin{eqnarray}
 \protect{\cal L}_q
  &=& \overline q (i\gamma^\mu \partial_\mu - m_q) q\theta_V - B_0\theta_V
 + g_\sigma^q \overline q q \sigma - g_\omega^q \overline q \gamma_\mu 
 q \omega^\mu - {1\over 2} m_{\sigma}^2 \sigma^2 + {1\over 2}
 m_{\omega}^2 \omega^2,
\end{eqnarray}
where $m_q$ is the current quark mass, $B_0$ is the bag constant 
in vacuum (and will remain fixed), $g_\sigma^q$ and $g_\omega^q$ 
are the corresponding quark and meson coupling constants, 
and $\theta_V$ is a step function which is one inside the bag volume
and vanishes  outside.

In MFA, the meson fields are treated as classical fields, 
and the quark field $q(x)$ inside the bag satisfies the equation of motion
\begin{equation}
[i\gamma^\mu\partial_\mu - (m_q - g_\sigma^q\overline{\sigma})-
g_\omega^q\overline{\omega} \gamma^0] q(x) = 0,
\end{equation}
where $\overline{\sigma}$ and $\overline{\omega}$ denote the constant 
mean values of the scalar and the time component of the vector field 
in symmetric nuclear matter. The normalized solution for the 
lowest state of the quark is given by\cite{Tony94,MIT} 
\begin{equation}
q(t,\vec{r}) =  {N_0\over\sqrt{4\pi}}  e^{-i\epsilon_q t/R} \left (
\begin{array}{c} j_0(x r/R) \\ 
i\beta_q \bfgreek {\sigma} \cdot\hat{\bf r} j_1(x r/R) \end{array} \right )
\, \theta(R-r)\, \chi_q \, , \label{cavity}
\end{equation}
where 
\begin{eqnarray}
\epsilon_q &=& \Omega_q + g_\omega^q\overline{\omega} R, \hspace{1cm}
\beta_q = \sqrt{\Omega_q-m_q^*R \over \Omega_q + m_q^*R}\, , \\
N_0^{-2} &=& 2R^3 j_0^2(x)[\Omega_q(\Omega_q-1)+ m_q^*R/2]/x^2\, ,
\end{eqnarray}
with $\Omega_q \equiv \sqrt{x^2 + (m_q^*R)^2}$, 
$m_q^* \equiv m_q - g_\sigma^q\overline{\sigma}$, $R$ the bag radius, 
and $\chi_q$ the quark Pauli spinor. The eigen-frequency, $x\, (x_0)$, 
of this lowest mode in medium (free space)
is determined by the boundary condition at the bag surface,
\begin{equation}
j_0(x) = \beta_q j_1(x).
\end{equation}
The form of the quark wave function in Eq.~(\ref{cavity}) 
is almost identical to that of the solution in free space. 
However the parameters in the expression
have been substantially modified by the surrounding nuclear medium. 
Note that as the value of $g^q_\sigma \overline{\sigma}$  
is usually larger than $m_q$, the quantity, $\beta_q$, 
becomes larger than unity, which means the lower component of the
Dirac spinor is enhanced. 
Thus the quarks in the nucleon embedded in the nuclear medium
are more relativistic than those in a  free nucleon. 

The mean values of the scalar ($\overline{\sigma}$)
and vector ($\overline{\omega}$) fields in symmetric nuclear matter 
are self-consistently determined by solving 
the following set of equations:
\begin{eqnarray}
\overline{\omega} &=& {g_\omega\rho\over m_\omega^2}, \label{omega}\\
\overline{\sigma} &=& {g_\sigma\over m_\sigma^2} 
C(\overline{\sigma}) \rho_s 
= {g_\sigma\over m_\sigma^2} 
C(\overline{\sigma}){4\over (2\pi)^3}
\int^{k_F} d^3 k {m_N^*(\overline{\sigma})\over 
\sqrt{m_N^{*2}(\overline{\sigma}) + k^2}}, \label{scc}\\
m_N^*(\overline{\sigma}) &=& {3\Omega_q (\overline{\sigma}) \over R}
- { z_0\over R} + {4\over 3}\pi R^3 B_0, \label{mass}\\
{\partial m_N^*(\overline{\sigma}) \over \partial R} &=& 0, \label{equil}
\end{eqnarray}
where $\rho$ ($\rho_s$) is the baryon (scalar) density and 
$k_F$ is the nucleon Fermi momentum, $g_\sigma= 3g_\sigma^qS(0)$,
$g_\omega= 3g_\omega^q$, and 
the quantity, $C(\overline{\sigma})$, is defined by 
\begin{equation}
C(\overline{\sigma}) \equiv S(\overline{\sigma})/S(0)=
-\left({\partial m_N^*(\overline{\sigma})
\over\partial\overline{\sigma}}\right)/g_\sigma,
\end{equation}
with $S(\overline{\sigma}) = \int_{\rm bag} d^3r\, 
\overline{q}(\vec{r})q(\vec{r})$.
Using the  quark wave function for the MIT bag, Eq.(\ref{cavity}),  
$S(\overline{\sigma})$ can be explicitly evaluated: 
\begin{equation}
S(\overline{\sigma}) = 
[\Omega/2+m_q^*R(\Omega-1)]/[\Omega(\Omega-1)+m_q^*R/2]. \label{ssigma}
\end{equation}
The second term in Eq.(\ref{mass}), ($-z_0/R$), parametrizes the sum of the 
center-of-mass (c.m.) motion and gluon corrections\cite{Guichon96}.
It will be important to note that this center-of-mass treatment 
is different from that of Jin and Jennings~\cite{JJ96} 
when we consider later the reduction of the bag constant, $B$.

For a free nucleon, the quantities, $z_0$ and $B_0$, are
determined simultaneously by requiring
the free nucleon  mass to be $m_N(\rho=0) = 939 $ MeV  and by
the stability condition, Eq.(\ref{equil}),
for a given bag radius, $R_0$, in free space 
(treated as an input parameter).

It is worthwhile to note that the self-consistency condition in QMC is 
identical to that in QHD, except that in QHD one has 
$C(\overline{\sigma})=1$
\cite{Tony94}, which corresponds to a point-like nucleon.
All information on the internal structure of the nucleon is contained
in $C(\overline{\sigma})$.

Equivalently, the equation for the mean field, $\overline{\sigma}$,
can also be derived by the 
thermodynamic condition
$\left.{\partial E_{\rm total} \over 
\partial \overline{\sigma}}\right|_{R,\rho} = 0$,
where the total energy per nucleon, $E_{\rm total}/A$, 
for symmetric nuclear matter is given by
\begin{equation}
\frac{E_{\rm total}}{A} = 
{4\over (2\pi)^3\rho}\int^{k_F}\! d^3k\sqrt{{m_N^*}^2(\overline{\sigma})+ k^2} 
+ {m_\sigma^2\over 2\rho}\overline{\sigma}^2 
+ {g_\omega^2\over 2m_\omega^2}\rho .
\end{equation}

In the following, we will consider the modified QMC model 
which was proposed by Jin and Jennings\cite{JJ96}, 
and includes the effect of the reduction in 
the bag constant, $B$, (instead of a fixed value, $B_0$) 
in the nuclear medium. 

The main difference of the approach used in the present work from that of 
Jin and Jennings\cite{JJ96} (see also Fleck et al.~\cite{Yazaki90} and 
our earlier work) is that we do not compensate for 
the spurious center-of-mass motion by subtracting 
the average value of the momentum squared of the three quarks.
It has been shown that this subtraction, which gives a strong field 
dependence, would cause significant reduction of the vector potential. 
Actual calculations using a relativistic oscillator
in an external field, where the spurious center-of-mass motion can be
removed exactly,  show that this field dependence is quite small and thus
the simple subtraction procedure is now believed to be 
inappropriate\cite{Guichon96}.

When a nucleon is put into the nuclear medium, the bag as a whole reacts to
the new environment, thus both $z_0$ and $B$ may be modified somehow.
There has been some consideration of how $z_0$ might vary 
depending on the type of baryons\cite{DJ80} and
on the density\cite{BM96}.
However, in the present work, we assume that $z_0$ is a fixed constant.
In the MIT bag, the bag constant, $B_0$, is introduced to satisfy 
energy-momentum conservation and provides the necessary pressure 
to confine the quarks. 
It enters the nonlinear boundary condition,
$-{1\over 2}n_\mu \partial^\mu [\overline{q}(x)q(x)] = B_0$, of the model, 
where $n_\mu$ is a unit four vector normal to the bag surface.
In free space, the bag constant term contributes approximately $200 - 300$ MeV
to the nucleon energy. However, it is expected to be reduced 
at finite density and temperature. At a sufficiently high density, 
once nucleons overlap severely, the whole concept
of the bag is probably no longer realistic.
One would expect that deconfinement of quarks could occur, 
and the bag pressure might eventually become zero\cite{Bender97}.
Then one can expect that a decrease of the bag pressure should lead to 
a larger bag radius and hence larger r.m.s. 
electromagnetic radii. On the other hand, these have been 
constrained by experiment~\cite{sick,Day90,coul,deForest83}.

In our previous work\cite{medium97} 
we kept $z_0$ and $B_0$ fixed and used Eq.~(\ref{equil}) to determine 
the in-medium bag radius, $R$, for a finite nuclear density, $\rho$. 
Typically the quark r.m.s. radius, $r^*_q$,  
calculated by the bag wave function is slightly increased,  
although the bag radius, $R$, typically decreases by a few percent 
at normal nuclear density. 
The extent of nucleon swelling in QMC, with a constant value 
for the bag pressure, $B_0$, has been shown to be consistent with 
the experimental constraints\cite{medium97}.

Two models for the reduction in the bag constant, motivated by the work 
of Jin and Jennings\cite{JJ96}, are examined in the following.
In the direct coupling model, the bag constant is related to the 
mean scalar field, through
\begin{equation}
{B\over B_0} = \left(1 - g_\sigma^B {4\over\delta}
{\overline{\sigma}\over m_N}\right)^\delta, \label{B1}
\end{equation}
where $g_\sigma^B$ and $\delta$ are real, positive parameters.
The case, $\delta=1$, corresponds to the model studied by 
Blunden and Miller\cite{BM96}.
In the limit, $\delta\rightarrow\infty$, it becomes a simple 
exponential form, ${B/B_0}=e^{-4g_\sigma^B\overline{\sigma}/m_N}$.
On the other hand, in the scaling model, the bag constant is connected 
to the effective nucleon mass in the medium,
\begin{equation}
{B\over B_0} = \left({m_N^* \over m_N}\right)^\kappa, \label{B2}
\end{equation}
where $\kappa$ is a real, positive parameter. Note that the 
standard treatment of the QMC model can be recovered
in the limits, $g_\sigma^B =0$, for the direct coupling model 
and, $\kappa = 0$, for the scaling model.

Because of the additional $\overline{\sigma}$ dependence arising in 
Eq.(\ref{mass}), through $B (\overline{\sigma})$, the 
self-consistency condition for the scalar field has to be modified.
By a straightforward evaluation, 
we may replace  
the quantity, $C(\overline{\sigma})$,  in Eq.(\ref{scc}) by,   
\begin{equation}
C(\overline{\sigma})\to C_{DC}(\overline{\sigma}) \equiv
C(\overline{\sigma}) + {16\over 3g_\sigma}g_\sigma^B\pi R^3 
{B\over m_N^*}\left( 1-{4\over\delta}{g_\sigma^B\overline{\sigma}\over m_N^*}
\right)^{-1},
\end{equation}
for the direct coupling model, and 
\begin{equation}
C(\overline{\sigma})\to C_{SC}(\overline{\sigma}) \equiv
C(\overline{\sigma})
\left( 1-{4\over 3}\pi R^3{\kappa B\over m_N^*}\right)^{-1},
\end{equation}
for the scaling model. 
Clearly both models give rise to a smaller bag constant.
The values for both quantities, $C_{DC}(\overline{\sigma})$ and 
$C_{SC}(\overline{\sigma})$, decrease as the density 
(the value of the $\overline{\sigma}$) increases, which is a similar feature 
to that of the standard QMC model.

\section{electromagnetic radius of the nucleon}

The evidence concerning a possible nucleon size change in the nuclear medium
comes largely from lepton-nucleus scattering, in particular, from
quasielastic electron-nucleus scattering. 
At energy and momentum transfers of several hundred MeV, 
the virtual photon is able to ``see'' single nucleons. 
Analyses of the spectrum for the $(e,e'N)$ process at the quasielastic peak
allow one to extract information for a bound nucleon.
The most outstanding phenomena are perhaps the missing strength 
in the Coulomb sum rule\cite{coul} and 
the remarkable $y$-scaling behaviour\cite{Day90}. 
However, different models often lead to quite different interpretations.
Calculations with a semiclassical RPA theory suggested that the nucleon
form factors had to be modified, with the nucleon r.m.s. radii
inceased by about 13\% in $^{12}C$, 23\% in $^{40}Ca$ 
and 21\% in $^{56}Fe$\cite{Alberico87}. 
This swelling of the proton may  be interpreted as a distortion of 
the pionic cloud\cite{Ericson86} or as a reduction of the 
effective masses of mesons and nucleons\cite{Cheon92}.

The most convincing constraint on the increase of the nucleon r.m.s.
radius  is from the systematic analysis of experimental data covering 
different momentum transfer regions, where an appealing scaling
behaviour of the cross section clearly exists. 
Analogous to the Bjorken scaling in electron-nucleon deep-inelastic 
scattering, this so-called $y$-scaling in quasielastic electron-nucleus 
scattering reveals the nucleon momentum distribution in the nucleus.
The scaling feature makes this reaction an ideal testing ground 
for medium effects on the nucleon form factors.
For  recent comprehensive reviews on this subject 
we refer to Ref.~\cite{Day90}.
  
The idea of $y$-scaling is based on an analysis 
under the following basic assumptions\cite{West75}: 
i) impulse approximation (single nucleon knockout), 
ii) no final state interactions, and 
iii) nonrelativistic nucleon kinematics.
These assumptions can be more or less justified by the careful selection of 
a certain kinematic region. 
At energies between the giant resonances and pion production
threshold for intermediate three momentum transfers 
(e.g., larger than 500 MeV ) 
inclusive electron-nucleus  scattering is dominated 
by the quasielastic peak, which results from scattering 
incoherently from individual nucleons embedded in the medium.
In the laboratory frame, the differential cross section
for this process in the plane-wave impluse approximation can 
be expressed as\cite{Rome,Sick85}
%
\begin{eqnarray}
{d^2\sigma\over d\Omega_e d\nu} =  \sum_i^A\int\!d^3k\int\!dE \,
  &\sigma_{ei}&(\vec{q},\omega;\vec{k},E)  S_i(\vec{k},E) \nonumber\\
  &&\delta( \nu+M_A-\sqrt{m_N^2+\vec{k'}^2}-\sqrt{M_{A-1}^2+\vec{k}^2} ),
\end{eqnarray}
where $\vec{k}$ and $\vec{k'}=\vec{k}+\vec{q}$
 are the initial and final momenta 
of the struck nucleon,
$E$ is the energy of the initial nucleon, and 
$(\nu, \vec{q})$ is the four-momentum of the virtual photon. 
$S(\vec{k},E)$ is the spectral function (the probability of finding a nucleon
of momentum $\vec{k}$ and energy $E$ in the nucleus).
$\sigma_{ei}$ is the elementary electron scattering
cross section from an off-shell nucleon. 
$M_A$ and $M_{A-1}$ are the masses of the target and residual nucleus.  

For sufficiently large $\vec{q}$, the cross section can be 
rewritten as
\begin{equation}
{d^2\sigma\over d\Omega_e d\nu} \simeq \overline{\sigma}_{e}(q,y)\cdot F(y),
\label{yscaling}
\end{equation}
where $q = |\vec{q}|$
and the scaling variable,
\begin{equation}
y = \left[ (M_A+\nu) \sqrt{\Lambda^2 - M^2_{A-1} W^2} - q \Lambda 
\right]/W^2,
\end{equation}
represents the momentum component of the nucleon parallel to $\vec{q}$ 
before the reaction.
$W \equiv \sqrt{ (M_A+\nu)^2 - q^2 }$ is the center-of-mass energy and
$\Lambda \equiv (M^2_{A-1} - m_N^2 + W^2)/2$.
The scaling function, $F(y)$, is  the probability to find a nucleon with
momentum fraction $y$ and is a universal  function
for different combinations of $q$ and $\nu$.
The quantity, $\overline{\sigma}_{e}(q,y)$, 
relates to the elementary electron  nucleon 
cross section.

Note that $y$-scaling is only valid within a well defined kinematic range. 
The sensitivity of scaling behaviour to the nucleon size can provide 
a reliable constraint on the possible nucleon size change in the medium. 
Any residual $q^2$ dependence could in principle be attributed to  
modified electromagnetic form factors of the nucleon. 
In this work, we do not try to reproduce the quasi-elastic electron  
scattering data, instead, we use the best estimate of the possible
change of size of a  bound nucleon available from $y$-scaling data 
to restrict the important model constant ---the bag constant, $B$.

Although little is known about the off-shell $\gamma NN$ vertex,
the prescription of de Forest\cite{deForest83} is believed 
to be successful and is valid at about 5\% near the quasielastic peak.  
By ascribing the off-shell correction to a kinematic factor, 
the single nucleon response is taken to be on-mass shell. 
More general discussions of this subject can be found in Ref.~\cite{Naus87}.
In the rest of this section, we calculate the electromagnetic
form  factors for an on-shell nucleon with the parameters determined
in the previous section in a self-consistent fashion.
To make the predictions as reliable as possible, we use the technique 
developed in Ref.~\cite{Lu97}, where corrections for spurious
center--of--mass motion and relativistic kinematics were considered.

As is well known, the electric ($G_E$) and magnetic ($G_M$)
form factors for an on-shell nucleon can be most conveniently 
evaluated in the Breit frame,
\begin{eqnarray}
\bra {N_{s'}({\vec{q}\over 2}) }  J^0(0) \ket {N_s(-{\vec{q}\over 2})} 
&=& \chi_{s'}^\dagger\,\chi_s\, G_E(Q^2), \label{GE} \\
\bra {N_{s'}({\vec{q}\over 2}) } \vec{J}(0) \ket {N_s(-{\vec{q}\over 2})} &=& 
\chi_{s'}^\dagger\, {i\bfgreek\sigma\times\vec{q}\over 2 m_N^*}\,\chi_s\, 
G_M(Q^2), \label{GM}
\end{eqnarray}
where $Q^2 \equiv -q^2 = \vec{q}^{\,2}$, $\chi_s$ and $\chi^{\dagger}_{s'}$ 
are Pauli spinors for the initial and final nucleons respectively.
The major advantage of the Breit frame is that $G_E$ and $G_M$
are explicitly  decoupled and can be determined separately 
by the time and space components of the electromagnetic current ($J^{\mu}$).
Note that, in the above definitions, Eqs.~(\ref{GE}) and (\ref{GM}), 
both the initial and final states 
are physical states which incorporate meson clouds.

The electromagnetic current of the quark is simply
\begin{eqnarray}
j^\mu_q(x) &=& \sum_f Q_f e \overline{q}_f(x) \gamma^\mu q_f(x),
\label{current}
\end{eqnarray}
where $q_f(x)$ is the quark field operator 
for the flavor $f$ and $Q_f$ is its charge in units of proton charge $e$.
The momentum eigenstate of a baryon is constructed 
by the Peierls-Thouless (PT) projection method\cite{Lu97,PT62},
\begin{equation}
\Psi_{\rm{PT}}(\vec{x}_1, \vec{x}_2, \vec{x}_3; \vec{p}) = 
N_{\rm{PT}} e^{i\vec{p} \cdot \vec{x}_{\rm{c.m.}} }
q(\vec{x}_1 - \vec{x}_{\rm{c.m.}}) q(\vec{x}_2 - \vec{x}_{\rm{c.m.}})
q(\vec{x}_3 - \vec{x}_{\rm{c.m.}}), \label{PTWF}
\end{equation}
where $N_{\rm{PT}}$ is a normalization  constant, 
$\vec{p}$  the total momentum of the baryon, and 
$\vec{x}_{\rm{c.m.}} = (\vec{x}_1 + \vec{x}_2 + \vec{x}_3)/3$ 
is the center of mass of the baryon (we assume equal mass quarks here).
It can be shown that the PT wave function satisfies the condition of
translational invariance.
Using Eqs.~(\ref{current}) and (\ref{PTWF}), the nucleon electromagnetic
form factors for the proton's quark core can be easily calculated by
\begin{eqnarray}
G_E(Q^2) &=& \int\! d^3r j_0(Qr)\rho_q(r)K(r)/D_{\rm PT}, \label{PTE}\\
G_M(Q^2) &=& (2 m_N^*/Q)\int\! d^3r j_1(Qr) 
\beta_q j_0(x r/R)j_1(x r/R)K(r)/D_{\rm PT}, \label{PTM}\\
D_{\rm PT} &=& \int\! d^3r \rho_q(r) K(r),
\end{eqnarray}
where 
$D_{\rm PT}$ is the normalization factor,
$\rho_q(r) \equiv j_0^2(x r/R) + \beta_q^2 j_1^2(x r/R)$, and 
$K(r) \equiv \int\! d^3u \, \rho_q(\vec{u}) \rho_q(-\vec{u} - \vec{r})$
is the recoil function to account for the correlation of the 
two spectator quarks.

At this stage, there is no satisfactory covariant treatment for the MIT bag 
model. On the other hand, relativistic effects are important for
most dynamic variables, 
especially for form factors at large momentum transfer.
They lead to a sizable correction for the r.m.s. radius of the nucleon.
In this paper,  we use a semi-phenomenological method to account for the 
relativistic corrections which is consistent with a mean field treatment of 
the QMC model.
Since a static MIT bag is an extended spherical object, it would be deformed 
if it were viewed in a moving frame of reference.
It is crucial to include
this Lorentz contraction of the bag for calculating form factors 
at moderate momentum transfer\cite{Lu97,LP70}.
In the prefered Breit frame, the photon-quark interaction can be 
reasonably treated as an instantaneous interaction. 
The resulting form
factors can be expressed through a simple rescaling, i.e.,
\begin{eqnarray}
G_E(Q^2) &=& ({m_N^*\over E^*})^2 G^{\rm sph}_E(Q^2 {m_N^*}^2/{E^*}^2), \\
G_M(Q^2) &=& ({m_N^*\over E^*})^2 G^{\rm sph}_M(Q^2 {m_N^*}^2/{E^*}^2),
\end{eqnarray}
where $E^*=\sqrt{{m_N^*}^2 + Q^2/4}$ and 
$G_{M,E}^{\rm sph}(Q^2)$ are the form factors calculated 
with the  spherical bag wave function [Eqs.~(\ref{PTE}) and (\ref{PTM})].
The scaling factor in the argument arises from the coordinate transformation
of the struck quark whereas
the prefactor, $(m_N^*/E^*)^2$,  comes from  the reduction 
of the integral measure of two spectator quarks in the Breit frame\cite{LP70}.
This prefactor is not unambiguously determined, but it is far less 
important in practice than the Lorentz contraction which rescales $Q^2$. 
The corresponding electromagnetic radii squared are thus given by,
\begin{equation}
r^2_{E,M} \equiv \left< r^2 \right>_{E,M} = -6\left. 
{dG_{E,M}(Q^2)\over d  Q^2}
\right|_{Q^2\rightarrow 0}.
\end{equation}

\section{pion clouds and meson exchange currents}

Before we proceed further it is useful to review the role of meson clouds.
As is well-known, a realistic picture of the free nucleon should include
the surrounding meson clouds.
Following the cloudy bag model (CBM)\cite{CBM,TT83},
we limit our considerations to the meson cloud corrections from the 
most important component,  namely the pion cloud.
In free space, the pion is a Goldstone boson in the chiral limit and 
is necessary to maintain chiral symmetry. In an infinite nuclear medium, 
the mean value  of the pion field vanishes due to parity conservation, 
however, the virtual process of pion exchange is always possible. 

In the QMC model, the medium effect on the nucleon is induced by 
the $\sigma$ and $\omega$ meson fields. The solution for a bound nucleon 
has already incorporated certain many body correlations due to 
the self-consistent solution of the equations of motion.
Estimates show that the contributions of meson exchange currents (MEC) to the 
quasielastic cross section do not show the $y$-scaling feature, and are thus 
expected to be irrelevant for the present study~\cite{MEC}.
One of the major implications of the $y$-scaling analysis 
of quasielastic scattering data is that MEC corrections appear 
to contribute primarily to the behaviour of the nucleon momentum 
distribution, rather than to a change of the effective electromagnetic 
form factors of the nucleon. Hence we 
consider it is reasonable to neglect possible explicit 
two- or more-body MEC contributions for the present study. 
Thus, we limit the meson cloud corrections
to one-pion loop on the same nucleon in the usual spirit of the CBM.

The Lagrangian related to the pion field and its interaction 
(we use  quark-pion pseudoscalar coupling),
up to leading order of $1/f_\pi$, is\cite{CBM}
\begin{equation}
\protect{\cal L}_{\pi q} = 
 {1\over 2} (\partial_\mu \bfgreek{\pi})^2
        - {1\over 2} m^2_\pi \bfgreek{\pi}^2
        - {i\over 2f_\pi} \overline q \gamma_5 \bfgreek{\tau} \cdot
        \bfgreek{\pi} q \delta_S, 
\end{equation}
where $\delta_S$ is  a surface delta function of the bag,
$m_\pi$  the pion mass and $f_\pi$  the pion decay constant.
The electromagnetic current of the pion is 
\begin{equation}
j^\mu_\pi(x) = -i e [ \pi^\dagger(x) \partial^\mu \pi(x)
               -\pi(x) \partial^\mu \pi^\dagger(x)],
\end{equation}
where
$ \pi(x) = {1\over \sqrt{2}}[\pi_1(x) + i\pi_2(x)]$
 either destroys a negatively charged pion
or creates a positively charged one.
As long as the bag radius is above 0.7 fm, 
the pion field is relatively weak and can be treated perturbatively\cite{CBM}. 
A physical baryon state, $\ket{A}$,  can be expressed as\cite{TT83}
\begin{equation}
\ket A \simeq \sqrt{Z_2^A} [ 1 + (m_A - H_0 - \Lambda H_I \Lambda )^{-1} H_I ] 
\ket {A_0} \label{state},
\end{equation}
where $\Lambda$ is a projection operator which projects out all the components
of $\ket A $ with at least one pion, $H_I$ is the interaction Hamiltonian
which describes the process of emission and absorption of pions.
The matrix elements of $H_I$ between the bare baryon states and 
their properties are\cite{TT83}
\begin{eqnarray}
v^{AD}_{0j}(\vec{k}) &\equiv& 
\bra {A_0} H_I \ket{{\bf \pi}_j(\vec{k}) D_0} = {i f^{AD}_0\over m_\pi}
{u(kR) \over [2\omega_k (2\pi)^3]^{1/2}} \sum_{m,n}
C^{s_D m s_A}_{S_D 1 S_A} (\hat{s}^*_m \cdot {\vec k}) 
C^{t_D n t_A}_{T_D 1 T_A} (\hat{t}^*_n \cdot {\vec e}_j),\\
w^{AD}_{0j}(\vec{k}) &\equiv& 
\bra{A_0 {\bf \pi}_j(\vec{k})} H_I \ket {D_0}
 = \left[v^{DA}_{0j}(\vec{k})\right]^* 
= -v^{AD}_{0j}(\vec{k}) = v^{AD}_{0j}(-\vec{k}),
\end{eqnarray}
where the pion has momentum $\vec{k}$ and  isospin projection $j$,
$f_0^{AD}$ is the reduced matrix element for the 
$\pi D_0 \rightarrow A_0$ transition vertex, $u(kR) = 3j_1(kR)/kR $,
$\omega_k = \sqrt{k^2+ m^2_\pi}$, and $\hat{s}_m$ and $\hat{t}_n$ 
are spherical unit vectors for spin and isospin, respectively.
Note that the $\pi NN$ form factor $u(kR)$ is fully determined 
by the model itself and  depends only on the bag radius. 
The bare baryon probability in the physical baryon state, $Z^A_2$, 
determined by the normalization condition, is
\begin{equation}
Z^A_2 = \left[ 1 +
 \sum_D \left({f^{AD}_0\over m_\pi}\right)^2 {1\over 12\pi^2}
\, \mbox{P}\!\int_0^\infty\! {dk\, k^4 u^2(k R)\over 
\omega_k (m_A - m_D - \omega_k)^2}
\right]^{-1},
\end{equation}
where P denotes a principal value integral.

For the detailed expressions for the pion loop  contributions, 
we refer to Ref.~\cite{Lu97}. The only modifications one should consider, 
for the purposes of studying nuclear matter, 
are to replace the bare masses and coupling constant with  those in medium,
i.e., to make the substitutions: 
$m_\pi \rightarrow m_\pi^*$, $m_D   \rightarrow m_D^*$ 
and $f_{\pi AD} \rightarrow f_{\pi AD}^*$.
In principle, the existence of the $\pi$ and $\Delta$ inside the nuclear 
medium would change the mean fields and 
probably lead to some modification of their properties.
Since the pion is well approximated as a Goldstone boson, 
the explicit chiral symmetry breaking is small in free space, 
and it should be somewhat smaller in nuclear medium\cite{BR91}. 
While the effective pion mass, $m_\pi^*$, may be slightly changed in the 
medium~\cite{waas}, we still use $m_\pi^*= m_\pi$, because the pion field 
contributes relatively little to the form factors 
(other than the neutron charge form factor, $G_{\rm En}$). 
As the $\Delta$ is treated on the same footing as the nucleon in the CBM,
its mass should vary in a similar manner to that of the nucleon. Thus
we assume that the in-medium and free space $N-\Delta$ mass splitting are
approximately equal, i.e., 
$m_\Delta^* - m_N^* \simeq m_\Delta - m_N$. This approximation is exactly 
satisfied in the standard QMC model (See Ref.~\cite{Tony94}).
In free space, the physical $\pi AD$ coupling constant is obtained by
$ f^{AD} \simeq \left({f^{AD}_0\over f^{NN}_0}\right) f^{NN}$.
There are corrections to the bare coupling constant, $f^{NN}_0$,
such as those from the nonzero quark mass and the correction from 
spurious center of mass motion.
Therefore, following the standard treatment, in practice we use the 
renormalized coupling constant in our calculation here, $f^{NN} \simeq 3.03$, 
which corresponds to the usual $\pi NN$ coupling constant, 
$f^2_{\pi NN}\simeq 0.081$. In the medium, the $\pi NN$ coupling constant
might be expected to decrease  slightly due to the enhancement of the 
lower component of the quark wave function, but we shall
ignore this small density dependence in the present study 
and use the approximate relation, $f_{\pi NN}^* \simeq f_{\pi NN}$, 
for simplicity.

\section{Discussion of the Results}

\begin{figure}
\vspace{2.5cm}
\centering{\
\epsfig{file=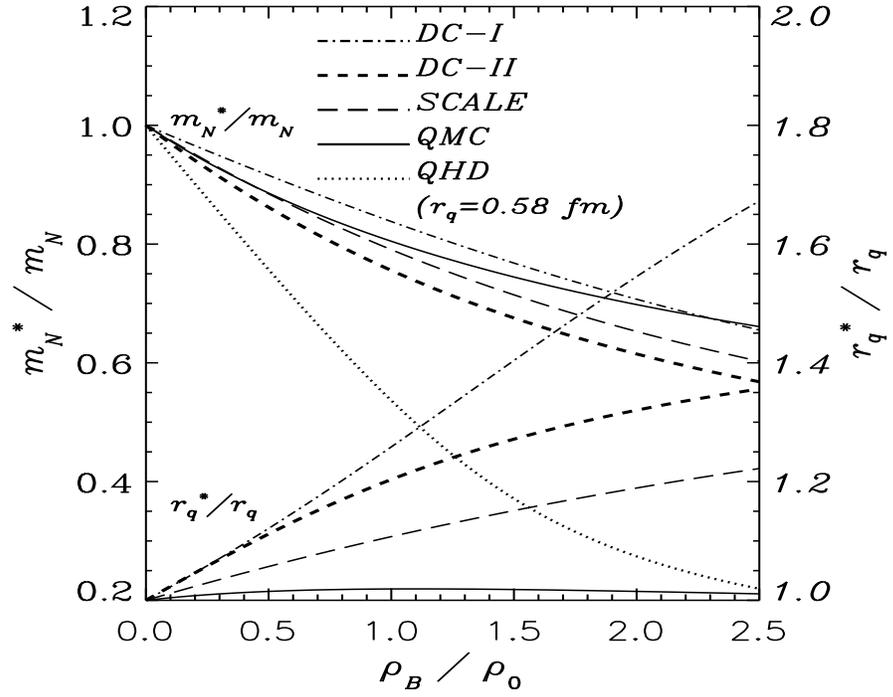,height=10cm,width=12cm}
\vspace{1cm}
\caption{Density dependence of the effective nucleon mass and 
quark r.m.s. radius. The parameters used in the calculation 
are given in Table~\protect\ref{QMC}.}
\label{mass.ps}}
\end{figure}

\begin{figure}
\centering{\
\epsfig{file=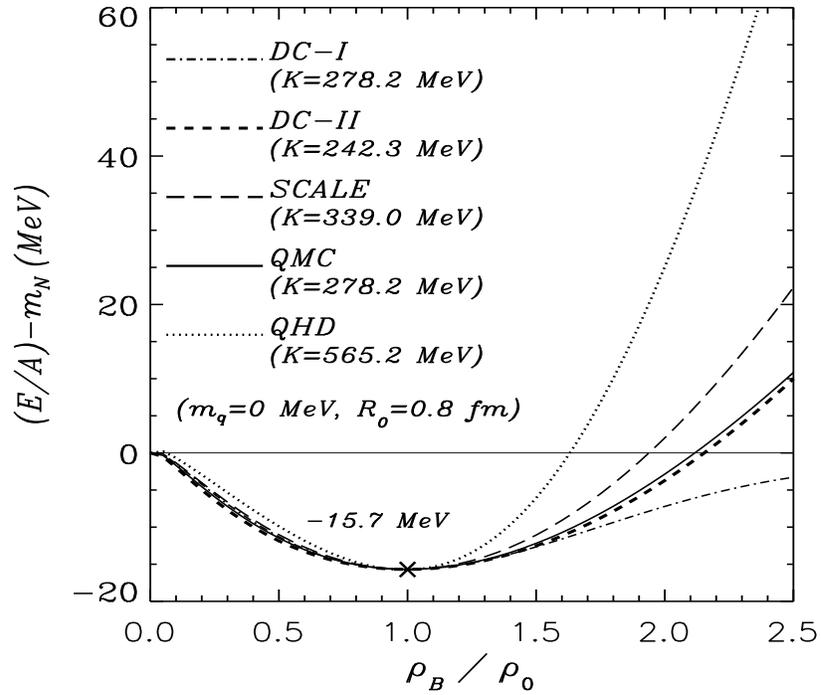,height=10cm,width=12cm}
\caption{Calculated values of the energy per nucleon (with the free nucleon 
mass subtracted) for symmetric nuclear matter. The values, $K$, 
denote the nuclear incompressibility calculated in each model.}
\label{saturat.ps}}
\end{figure}

\begin{figure}
\centering{\
\epsfig{file=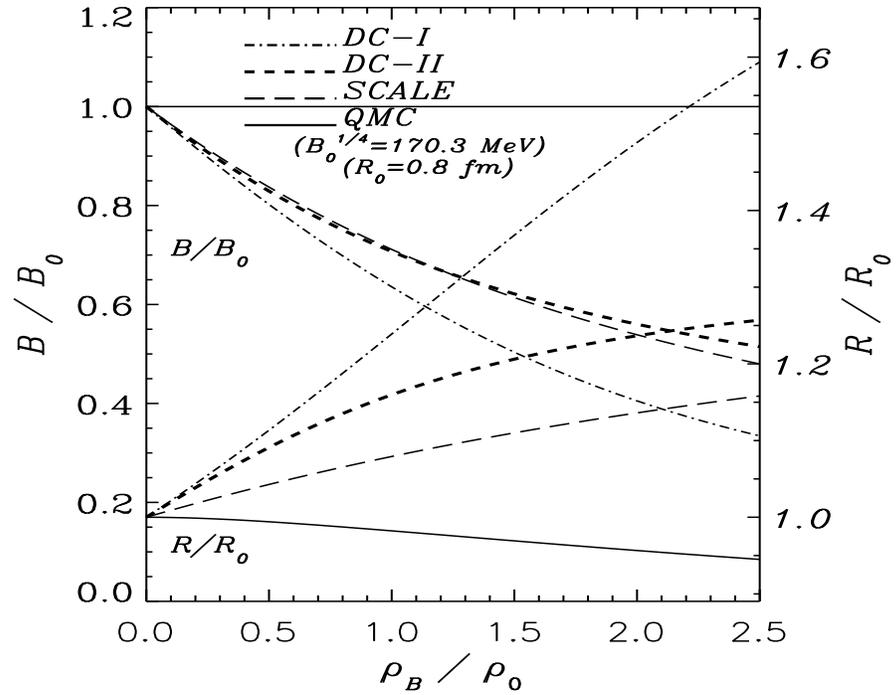,height=10cm,width=12cm}
\caption{Density dependence of the bag constant and the bag radius.} 
\label{bagconst.ps}}
\end{figure}

In the previous standard (i.e., ``constant $B_0$'') QMC models,  
typical solutions show that the values of the scalar mean field are 
quite large -- the term $g_\sigma^q\overline{\sigma}$ is of the order 
of 100 MeV at normal nuclear density, and 
thus the final results are insensitive to the values used for the 
current quark mass, $m_q$~\cite{Tony94,Guichon96}.
Later variations introduced  some density dependence of the bag constant, $B$,
through a parametrized, phenomenological  form (static parameter $g_\sigma^B$ 
and $\delta$ or $\kappa$)\cite{JJ96}. 
Thus for a given bag radius (input for free nucleon), there are two 
truly dynamical free parameters, the quark and meson coupling constants, 
$g_\sigma^q$ and $g_\omega^q$, which are adjusted to reproduce 
the saturation properties of symmetric nuclear matter. 
The effective nucleon mass, the bag radius in matter, and the value for 
the scalar mean field are all determined by the equations of motion, 
whereas the mean field of the vector meson is directly related to the 
baryon density (see also Eqs.(\ref{omega}) -- (\ref{equil})). 
A typical set of parameters and the value of the nuclear 
incompressibility, $K$, obtained for each model -- 
direct coupling models I and II (DC-I and DC-II), the scaling model (SCALE) 
and the standard QMC --  are listed in Table~\ref{QMC}. 
The properties of these self-consistent solutions as a function of
the nuclear density (in units of normal nuclear matter density, 
$\rho_0 = 0.15$ fm$^{-3}$), are shown in 
Figs.~\ref{mass.ps}--\ref{bagconst.ps},  where the bag radius in free space 
is taken to be 0.8 fm. For a study of the (rather weak) dependence of 
the final results on the values chosen for the bag radius in free space, 
see Refs.~\cite{Tony94,Guichon96}.

In Fig.~\ref{mass.ps}, $r^*_q\, (r_q)$ denotes the quark r.m.s. radius 
calculated using the quark wave function 
in medium (free space), and $m^*_N\, (m_N)$ denotes the nucleon mass
in medium (free space). The density dependence of the 
effective nucleon mass is more or less similar for all 
the QMC type models considered in this study, 
and the reduction of the mass is very moderate 
compared with that of QHD.  
However, the QMC type models among themselves show a quite different 
density dependence of the quark r.m.s. radius, $r^*_q$.

In Fig.~\ref{saturat.ps}, we show the energy per nucleon 
(with respect to the free nucleon mass, $E/A - m_N$)  
calculated for each model, together with the the value of the nuclear 
incompressibility, $K$. The parameters in each model are fitted to 
reproduce the energy at the saturation point, $-15.7$ MeV.  
All the QMC type models reproduce reasonable values ($200 \sim 300$ MeV) 
for the nuclear incompressibility, $K$ (for the parameter sets listed in 
Table~\ref{QMC}). As is well known, the value calculated in a naive version 
of the QHD model is too large, and thus, the energy per nucleon grows more 
rapidly than those of the QMC type models as the density increases.

We also show the density dependence of the bag constant, $B$, 
and bag radius, $R$, in Fig.~\ref{bagconst.ps}.
Although the bag radius calculated for the standard QMC model decreases
slightly as the density increases, the increase of the bag radius   
calculated for the modified QMC models is very pronounced.
As expected, a large reduction of the bag constant causes 
a large increase of the bag radius.

The quality of the nucleon electromagnetic form factors in free space
has been described in our previous publication\cite{Lu97}.
The corrections for the center-of-mass motion and Lorentz contraction
lead to a significant improvement over earlier static bag model calculations, 
in particular, at moderate momentum transfers. 
The pionic correction tends to soften the both form factors, $G_E$ and $G_M$, 
and thus increase the corresponding electromagnetic r.m.s. radii 
of the proton, $r_E \equiv \left<r^2_E\right>^{1/2}$ and 
$r_M \equiv \left<r^2_M\right>^{1/2}$.  
(The corresponding quantities in free space will be denoted by 
$r^0_E$ and $r^0_M$, respectively.)
We also note that the size of these corrections is larger 
when smaller values of the bag radius in free space are chosen. 

As in our previous study\cite{medium97}, the charge form factors 
are much more sensitive to the nuclear medium than the magnetic ones. 
The variation of the magnetic form factors with respect to their 
free space counterparts is  nearly one order of 
magnitude smaller than that for the charge form factors.
As for the density dependence of the electromagnetic form factors,  
they are clearly suppressed at small $Q^2$, 
relative to those in free space. This indicates  that the 
electromagnetic radii of the bound nucleon have increased. 
In other words, the nucleon becomes ``swollen'' once it is put into 
the nuclear medium. For a fixed $Q^2$ (less than $0.3 \mbox{ GeV}^2$), 
the form factors decrease almost linearly with the increase 
in the nuclear density. 
At $Q^2 \sim 0.3 \mbox{ GeV}^2$, the proton and neutron  charge form factors 
are  reduced by roughly 5\% and 6\% at $\rho = 0.5 \rho_0$, and 
8\% at normal nuclear density, $\rho_0 = 0.15$ fm$^{-3}$. 
Similarly, the proton and neutron magnetic form factors are 1\% and 0.6\% 
smaller at $\rho = 0.5 \rho_0$,
and 1.5\% and 0.9\% at normal nuclear density.

 
As explained earlier, the best experimental constraints on the changes 
in these electromagnetic form factors
come from the analysis of $y$-scaling data.
In the kinematic range normally covered by this analysis, 
the electron-nucleon scattering cross section is predominantly magnetic, 
so that the limit extracted applies essentially to $G_{M}$. 
For example, in Fe it was concluded by Sick that the
nucleon r.m.s. radius cannot vary by more than 3\%\cite{sick}. 
For the standard QMC model, with a fixed bag constant, the calculated increase 
in the r.m.s. radius of the magnetic form factors is less than
0.8\% at $\rho_0$. This is well within the experimental limit.
For the modified QMC models (with smaller $B$), however, 
the variation of the r.m.s. radius can be much stronger. 

In Tables~\ref{B0.8}---\ref{scale1.0}, we show the changes of the  
electromagnetic r.m.s. radii of the proton,  
$\delta r_{E,M} \equiv (r_{E,M} - r^0_{E,M})/r^0_{E,M}$, vs
the bag constant for three different nuclear densities. Two values of the bag 
radius in free space are used in this calculation. 
In the direct coupling model, we take the  
limit, $\delta\rightarrow \infty$, and vary $g_\sigma^B$. 
For each selected model (fixed $g_\sigma^B$ or $\kappa$), the coupling
constants, $g_\sigma^2/4\pi$ and $g_\omega^2/4\pi$, are first determined 
by fitting 
the saturation properties of nuclear matter at normal nuclear density. 
The mean fields, quark frequency, and effective nucleon mass are then 
subsequently solved self-consistently for other nuclear densities.
The current quark mass is taken to be zero. 
It is clear from the Tables that the bag constant and the electromagnetic 
r.m.s. radii are strongly correlated. 
A smaller bag constant will lead to a larger size of the bound nucleon.

\begin{figure}
\centering{\
\epsfig{file=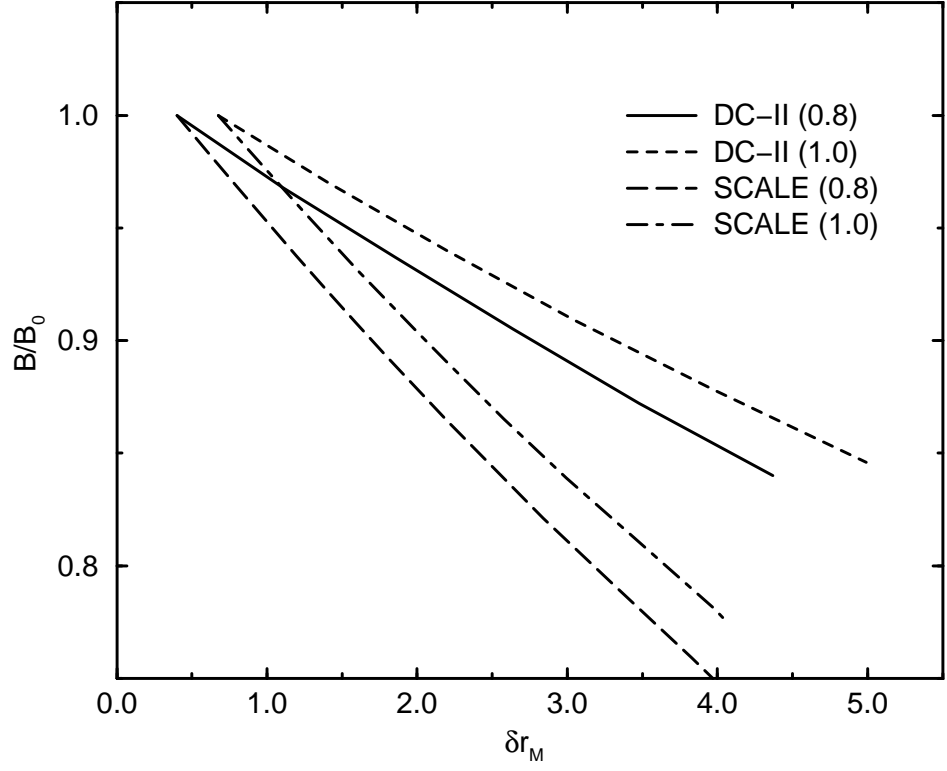,height=12cm}
\caption{Correlation of the bag constant with the magnetic radius of 
the proton at the average nuclear density, $\rho/\rho_0=0.7$. 
The number in the bracket refers to the bare bag radius, $R_0$.}
\label{rms.ps}}
\end{figure}

For finite nuclei, it is well known that a large portion of the nucleons 
are distributed near the nuclear surface. 
The experimental constraints thus correspond to the quantities 
at the average nuclear density, which we take to be 
about $0.7\rho_0$\footnote{Using standard nuclear density distributions
the average densities of $^{16}O$, $^{40}Ca$ and $^{56}Fe$ are $0.6\rho_0$,
$0.7\rho_0$ and $0.7\rho_0$, respectively.}.
Fig.~\ref{rms.ps} shows the correlation between the bag constant and the 
magnetic radius of the proton at a nuclear density $\rho=0.7\rho_0$.
For an increase of about 3\% in the magnetic radius 
($\delta r_M \simeq 3 \%$), which is  the upper limit allowed
from the $y$-scaling data, the bag constant is allowed to decrease by 
about 8\% for the direct coupling model and 16\% for the scaling model,
with the bare bag radius, $R_0=1.0$ fm.  
These reductions become 12\% and 18\% with a smaller bag radius, $R_0=0.8$ fm.


The possible reduction of the bag constant, $B$, might also be extracted 
from the electric form factors if we had a reliable experimental constraint
for the charge radius. 
In the $y$-scaling analysis, since the electric and magnetic form factors 
contribute typically in the ratio 1:3 in the kinematic regime selected,
the corresponding limit on $G_{E}$ would be nearly 10\%. 
For the electric form factors the best experimental limit seems to 
come from the Coulomb sum-rule, where a variation
bigger than 4\% would be excluded\cite{coul}. 
With this strong limit, the bag constant
is allowed to be reduced by roughly 10\% in the scaling model 
and is almost unchangeable in the direct-coupling model .


\section{Conclusion}

In summary, the bag constant in the nuclear medium, as a crude representation 
of the very complicated QCD mechanism of confinement, may be expected to
decrease. We have examined the possible variations of the bag constant
in two modified versions of the QMC model. The parameters are determined 
to reproduce the empirical values for the bulk properties of symmetric 
nuclear matter. We have calculated the density dependence of the 
electromagnetic form factors of the bound nucleon in the medium, 
then compared the corresponding r.m.s. radii with the best available 
$y$-scaling analysis for quasielastic electron-nucleus scattering.
Since the change in the size of the nucleon is strongly limited by 
these $y$-scaling data, so is the bag constant.
Our results show that the bag constant in medium, $B$, is allowed to 
decline by only 10--17\% at the typical average nuclear density of about
$0.7\rho_0$\footnote{In terms of the actual models studied this means that
$g_\sigma^B$ should be less than 1.2 in the direct coupling model 
with $\delta\rightarrow\infty$ (c.f. Eq.~(\ref{B1})) and 
$\kappa$ less than 1.2 in the scaling model (c.f. Eq.~(\ref{B2})).}. 
This limit would be reduced by a further 7\% at normal nuclear density. 
This is a considerably smaller change than most of the cases studied by
Jin and Jennings\cite{JJ96}, but comparable with the usual estimates of the
change in the gluon condensate\cite{gluon}, 
or of the bag constant itself\cite{Bender97} in nuclear matter.

It is worth mentioning that the pion cloud of the nucleon plays an 
important role in calculating the electromagnetic form factors. 
Without the pion cloud, the nucleon r.m.s. radius tends to be somewhat small,
for a reasonable bag radius. Thus the relative variation is much more
sensitive to the bag constant. Typically, $\delta r_M$ 
without the pionic correction is three times larger
than that with the pion field.
 We would like to emphasize that the MEC effects which can be important 
in conventional nuclear physics have not been included in the present study. 
This should be reasonable in a first treatment because MEC corrections 
are not very important 
near the quasielastic peak and they contribute primarily to the behaviour 
of the nucleon momentum distribution, rather than to a change of the 
effective electromagnetic form factors of the nucleon. 
Further improvements rely on a better treatment of the QMC model and 
require the explicit inclusion of nucleon-nucleon correlations 
and medium polarization effects.

We are pleased to thank I. Sick for helpful correspondence
concerning the constraints from $y$-scaling analysis.
We are also grateful to M. Ericson for informative discussions.
This work was supported by the Australian Research Council. 
A.W.T and K.S. acknowledge 
support from the Japan Society for the Promotion of Science.

\begin{table}
\caption{Saturation properties calculated in each model for symmetric 
nuclear matter at normal nuclear density, $\rho_0 = 0.15$ fm$^{-3}$. 
DC-I and DC-II refer to the direct coupling models with $g_\sigma^B$ = 4 and  
$\delta$ = 890, and $g_\sigma^B$ = 2.8 and $\delta\rightarrow\infty$, 
respectively. SCALE denotes the scaling model with $\kappa$ = 1.45 and 
QMC denotes the standard QMC model treatment. 
(The bag radius in free space is taken to be $R_0$ = 0.8 fm.)
$x\, (x_0)$ is the bag eigenfrequency in matter (vacuum), and $K$ is  
the nuclear incompressibility.}
\label{QMC}
\begin{center}
\begin{tabular}{l|ccccccr}
model& $g_\sigma^2/4\pi$ & $g_\omega^2/4\pi$ &$x/x_0$ 
& $m_N^*/m_N$ & $R/R_0$ & $B/B_0$ & $K(MeV)$ \\ \hline
DC-I   &  0.521& 1.600&  0.9393&  0.8380&   1.239& 0.636& 278.2\\
DC-II  &  2.811& 6.393&  0.8354&  0.7557&   1.160& 0.707& 242.3\\
SCALE  &  3.185& 8.525&  0.8941&  0.7907&   1.079& 0.711& 339.0\\ 
QMC    &  5.373& 5.257&  0.8368&  0.8048&   0.982& 1.000& 278.2\\
\end{tabular}
\end{center}
\end{table}

\begin{table}
\caption{Variation of the  bag constant vs the electromagnetic radii of 
the proton in the direct coupling model with the limit, 
$\delta\rightarrow\infty$, for $R_0$ = 0.8 fm. 
The calculated values of the 
electromagnetic r.m.s. radii in free space are, 
$r^0_E \equiv \left< r^2 \right>_{Ep}^{1/2}$ = 0.736 fm and 
$r^0_M \equiv \left< r^2 \right>_{Mp}^{1/2}$ = 0.731 fm. 
Below, $\delta r_{E,M} \equiv (r_{E,M} - r^0_{E,M})/r^0_{E,M}$ and are given 
as  percentages. 
}
\label{B0.8}
\begin{center}
\begin{tabular}{l|c|c|ccc|ccc|ccc}
 & & & \multicolumn{3}{c|}{$\rho/\rho_0=0.5$} 
 & \multicolumn{3}{c|}{$\rho/\rho_0=0.7$}
 & \multicolumn{3}{c}{$\rho/\rho_0=1.0$} \\ 
 $g_\sigma^B$ & $g_\sigma^2/4\pi$ & $g_\omega^2/4\pi$
 	      & $B/B_0$ & $\delta r_E$  & $\delta r_M$  
              & $B/B_0$ & $\delta r_E$  & $\delta r_M$  
              & $B/B_0$ & $\delta r_E$  & $\delta r_M$ \\ \hline
0.4& 5.015& 5.604&  0.976& 3.20& 0.82&  0.968& 4.26& 1.10&  0.958& 5.66& 1.51\\
0.8& 4.651& 5.906&  0.952& 3.86& 1.37&  0.937& 5.17& 1.85&  0.916& 6.93& 2.52\\
1.2& 4.293& 6.168&  0.928& 4.56& 1.95&  0.905& 6.16& 2.64&  0.874& 8.32& 3.61\\
1.6& 3.935& 6.373&  0.903& 5.30& 2.56&  0.872& 7.22& 3.48&  0.832& 9.84& 4.78\\
2.0& 3.575& 6.502&  0.879& 6.08& 3.19&  0.840& 8.35& 4.37&  0.790& 11.5& 6.03\\
\end{tabular}
\end{center}
\end{table}

\begin{table}
\caption{Same as in Table~\protect\ref{B0.8} except $R_0$ = 1.0 fm. 
The free space values are now, $r^0_E$ = 0.83 fm and $r^0_M$ = 0.80 fm.}
\label{B1.0}
\begin{center}
\begin{tabular}{l|c|c|ccc|ccc|ccc}
 & & & \multicolumn{3}{c|}{$\rho/\rho_0=0.5$} 
 & \multicolumn{3}{c|}{$\rho/\rho_0=0.7$}
 & \multicolumn{3}{c}{$\rho/\rho_0=1.0$} \\ 
 $g_\sigma^B$ & $g_\sigma^2/4\pi$ & $g_\omega^2/4\pi$
	      & $B/B_0$ & $\delta r_E$  & $\delta r_M$  
              & $B/B_0$ & $\delta r_E$  & $\delta r_M$  
              & $B/B_0$ & $\delta r_E$  & $\delta r_M$ \\ \hline
0.4& 4.692& 4.837&  0.977& 3.81& 1.07&  0.970& 4.99& 1.41&  0.960& 5.78& 1.16\\
0.8& 4.341& 5.140&  0.954& 4.60& 1.65&  0.940& 6.07& 2.20&  0.921& 7.95& 2.92\\
1.2& 4.001& 5.418&  0.931& 5.45& 2.28&  0.909& 7.24& 3.05&  0.880& 9.58& 4.10\\
1.6& 3.668& 5.657&  0.907& 6.35& 2.95&  0.878& 8.51& 3.98&  0.840& 11.4& 5.39\\
2.0& 3.340& 5.844&  0.883& 7.31& 3.66&  0.846& 9.89& 4.99&  0.799& 13.3& 6.81\\
\end{tabular}
\end{center}
\end{table}

\begin{table}
\caption{Variation of the bag constant vs the electromagnetic radii of the 
proton in the scaling model with $R_0$ = 0.8 fm.
The values of the electromagnetic r.m.s. radii in free space are,
$r^0_E$ = 0.74 fm and 
$r^0_M$ = 0.73 fm. See also the caption of Table~\protect\ref{B0.8}.}
\label{scale0.8}
\begin{center}
\begin{tabular}{l|c|c|ccc|ccc|ccc}
 & & & \multicolumn{3}{c|}{$\rho/\rho_0=0.5$} 
 & \multicolumn{3}{c|}{$\rho/\rho_0=0.7$}
 & \multicolumn{3}{c}{$\rho/\rho_0=1.0$} \\ 
 $\kappa$ & $g_\sigma^2/4\pi$ & $g_\omega^2/4\pi$
	      & $B/B_0$ & $\delta r_E$  & $\delta r_M$  
              & $B/B_0$ & $\delta r_E$  & $\delta r_M$  
              & $B/B_0$ & $\delta r_E$  & $\delta r_M$ \\ \hline
1.0& 3.732& 7.219&  0.888& 3.75& 1.73&  0.850& 4.70& 2.11&  0.800& 7.21& 3.43\\
1.2& 3.468& 7.727&  0.866& 4.02& 2.03&  0.821& 5.59& 2.84&  0.761& 7.88& 4.07\\
1.4& 3.241& 8.359&  0.843& 4.34& 2.35&  0.791& 6.08& 3.32&  0.721& 8.67& 4.79\\
1.6& 3.021& 9.058&  0.819& 4.67& 2.68&  0.759& 6.63& 3.83&  0.681& 9.56& 5.57\\
1.8& 2.652& 9.215&  0.805& 4.78& 2.89&  0.740& 6.81& 4.13&  0.654& 9.95& 6.06\\
\end{tabular}
\end{center}
\end{table}
\begin{table}
\caption{Same as in Table~\protect\ref{scale0.8} except $R_0$ = 1.0 fm. 
The values of the electromagnetic r.m.s. radii in free space are, 
$r^0_E$ = 0.83 fm and $r^0_M$ = 0.80 fm.}
\label{scale1.0}
\begin{center}
\begin{tabular}{l|c|c|ccc|ccc|ccc}
 & & & \multicolumn{3}{c|}{$\rho/\rho_0=0.5$} 
 & \multicolumn{3}{c|}{$\rho/\rho_0=0.7$}
 & \multicolumn{3}{c}{$\rho/\rho_0=1.0$} \\
 $\kappa$ & $g_\sigma^2/4\pi$ & $g_\omega^2/4\pi$
              & $B/B_0$ & $\delta r_E$  & $\delta r_M$  
              & $B/B_0$ & $\delta r_E$  & $\delta r_M$  
              & $B/B_0$ & $\delta r_E$  & $\delta r_M$ \\ \hline 
1.0& 3.431& 6.125&  0.899& 4.27& 1.87&  0.865& 5.80& 2.58&  0.822& 7.91& 3.59\\
1.2& 3.183& 6.565&  0.879& 4.56& 2.16&  0.839& 6.23& 2.99&  0.787& 8.58& 4.21\\
1.4& 2.952& 7.064&  0.859& 4.87& 2.46&  0.813& 6.71& 3.44&  0.752& 9.33& 4.88\\
1.6& 2.791& 7.814&  0.835& 5.31& 2.84&  0.781& 7.36& 3.98&  0.712&10.38& 5.73\\
1.67&2.639& 7.762&  0.832& 5.30& 2.87&  0.777& 7.35& 4.04&  0.705&10.40& 5.83\\
\end{tabular}
\end{center}
\end{table}


\newpage
\vspace*{1cm}

\newpage
\vspace*{1cm}

\newpage
\vspace*{1cm}


\end{document}